# Challenges and Opportunities in the Transition from 5G to 6G Networks

Yousef Tahboub
School of Computing
Montclair State University
Montclair, New Jersey
tahbouby1@montclair.edu

Krishna Sampada Bodige
School of Computing
Montclair State University
Montclair, New Jersey
bodigek1@montclair.edu

Ryan Tipping
School of Computing
Montclair State University
Montclair, New Jersey
tippingr1@montclair.edu

***Abstract*— The rapid adoption of 5G revolutionized connectivity worldwide, enabling faster data transfer along with the support of a variety of use cases. However, next-gen technologies demand higher performance, as the limitations of 5G become increasingly apparent. The 6G network transition is approaching but comes with unique challenges, particularly with spectrum management, security, and data transmission. The present research considers the most significant challenges that must be addressed for a smooth and successful 5G-6G transformation to occur. Based on case studies and academic literature, the research employs a comparative approach and statistical analysis to evaluate 5G limitations and assess the innovative potential of 6G. The envisioned contribution is the identification of the most significant technical, regulatory, and security vulnerabilities, along with proposed solutions. Through these alternatives, the research will help to identify 6G networks that are secure, resilient, and applicable globally for use and additional critical infrastructure support.**

***Keywords—5G, 6G, spectrum management, security, interoperability, next-generation networks, transition strategies.***

## I. INTRODUCTION

Transitions between generations of cellular communication technologies have consistently reshaped global connectivity. The significant transitions from 3G to 4G and then 4G to 5G brought users great advancements such as improvements in speed, reliability, capabilities, high-quality streaming, and much more. However, with every technological evolution, new challenges also arise. While 5G has transformed connectivity by enabling low latency, massive machine-type communication, and enhanced broadband experiences, it is beginning to reach its operational and theoretical limits as the demand for intelligent, real-time, and data-intensive applications continues to grow.

From the analog voice systems of 1G to the digital revolution to 2G, wireless systems have evolved rapidly. 3G introduced mobile internet access, while 4G enabled broadband-level speeds and video streaming. 5G, in turn, focused on ultra-reliability and low-latency communications. This progression showcases each generation and how they addressed prior limitations whilst simultaneously creating new demands, setting the stage for 6G.

The transition to 6G represents the future of global communications infrastructure, with 100 times higher data rates than 5G, and with seamless integration of artificial intelligence and operational capabilities to enable holographic and augmented reality uses. The global vision for sixth-generation wireless networks is increasingly shaped by international standardization bodies rather than isolated vendor-driven innovation. Specifically, the International Telecommunication Union (ITU) has proposed the IMT-2030 framework that provides a long-term roadmap to the 6G connectivity systems focusing on smart connectivity, integrated sensing and communication, sustainability, and full global coverage [36]. Such a framework places 6G not just as an extension of mobile broadband but as a generalized digital infrastructure of the most vital services in healthcare, transport, industry, and public security. To add to this view, Saad et al. comment that 6G should be built as a platform that closely combines communication, computation, control, and sensing, instead of a system that will optimize data throughput [50]. These international efforts demonstrate that the transition to 6G is both a technological and socio-technical transformation, requiring coordinated innovation across academia, industry, and regulatory organizations. However, these opportunities are accompanied by significant challenges related to spectrum management, cybersecurity, and energy efficiency. Governments, researchers, and telecommunication industries are now creating frameworks and strategies to make this transition effortless and secure to 6G.

As connectivity becomes more immersive and data-centric, 6G networks will not only raise technical challenges but also introduce new ethical and security concerns. The integration of artificial intelligence, autonomous systems, and the growing number of interconnected devices increases the attack surface for cyber threats, and make network vulnerabilities far more complex to detect and mitigate. Furthermore, the massive amount of real-time data generated by users and smart infrastructure will heighten issues of privacy, surveillance and data ownership, needing stricter global regulations and transparency. To address these concerns, this paper will employ a comparative analysis of current 5G infrastructures and projected 6G framework, supported by case studies and data evaluation of network performance and cybersecurity standards.

The paper reviews the major challenges and opportunities that have led to the transition of 5G to 6G due to the constraints of the existing 5G technologies and the potential of 6G innovation. Using case studies and comparison, the research examines some important technical, regulatory, and security issues that need to be resolved to establish a deployable and reliable 6G network globally. Finally, the development of 5G to 6G is not just a generational improvement, but a revolution in the interaction of humans, machines, and places. This study will add to a safe and efficient next generation of telecommunication systems that is globally available by emphasizing the possible strategies and innovations.

## II. PROBLEM STATEMENT

The wireless communication development has hit a critical stage with 5G networks being deployed around the world to facilitate the development of improved mobile broadband, massive internet of things and ultra reliable low-latency communications. Although 5G has shown significant improvement in the performance regarding throughput, latency, and the density of devices compared to the previous generations, various studies have shown that these functions are not adequate to meet the growing requirements of the next decade [1], [2], [4]. Existing 5G deployments reveal limitations related to fragmented spectrum allocation, scaling challenges in ultra-dense deployments, energy efficiency, and supporting pervasive artificial intelligence and massive autonomous systems [2], [3], [5]. Consequently, the research community in academia and industry has already started defining the 6G of wireless networks that should be able to achieve terabit per second data rates, microsecond latency, AI-native network control, and service continuity throughout the world based on integrated terrestrial and non-terrestrial systems, in addition to enhancing sustainability and operational efficiency [1], [3], [6], [7], [17]. Beyond academic research, formal industry and standards-driven efforts further demonstrate that 6G development is progressing from conceptual exploration toward early specification. The 3GPP technical report TR 38.913 indicates future requirements of performance and operational expectations of future radio access technologies beyond 5G-Advanced, with extreme data rates, ultra-low latency, and closer integration with non-terrestrial networks [37]. Likewise, the NGMN Alliance has established deterministic performance, global service continuity, and AI-native automation across heterogeneous infrastructures as the main reasons behind 6G adoption [39]. These initiatives underscore that even as existing 5G systems continue to be enhanced, current systems are structurally inadequate to support the requirements over the long-term and thus underscore the need to proceed with a generational redesign and not incremental optimization.

However, along with this, there will come a myriad of technological, architectural, operational, and security difficulties in the process of switching 5G to 6G. Achieving the desired performance levels requires the use of higher frequency spectrum including sub-terahertz, terahertz frequency bands, which suffer from severe propagation loss, atmospheric absorption, and the lack of scalable, low-power transceiver hardware [1, 4]. Various proposed technologies that could enable this, such as reconfigurable intelligent surfaces, ultra-massive MIMO, integrated photonic communication components, have been proposed, but most are still in the early stages of research, lacking full-scale empirical testing and deployment maturity [1,11,17].

In addition, the move to 6G demands redesigning of network architecture. Conventional centralized systems cannot be used in the dynamic computation, mobility management and real-time decision making that are needed by the emerging smart environments. The 6G frameworks proposed feature hierarchical, distributed, and AI-based architectures with cloud, edge, and device-level intelligence and combining terrestrial networks with high-altitude systems, low Earth orbit satellites, and unmanned aerial communication systems to provide seamless coverage and autonomous network optimization [6], [7], [8], [17].

New challenges related to model transparency, robustness to adversarial attacks, decision accountability, and unintended algorithmic bias [9], [10], [18] are brought about by the deeper embedding of AI in network control functions such as spectrum allocation, routing, mobility management, and anomaly detection. There is also a critical need for large-scale data sharing across distributed nodes for AI-enabled network management. This involves data governance, privacy protection, ownership, and trust issues in decentralized learning environments that remain largely unaddressed [10], [13], [19]. Federated learning and edge intelligence have been proposed to avoid centralized data exposure; however, they introduce new vulnerability points and enhance system complexity [5], [10]. The growing size of the interconnected IoT and self-driving systems increases the area of the cyber-attacks on the physical, link, and network levels [19], [20]. Despite these prospects, new security models such as authentication systems built upon blockchains, post-quantum encryption, and adaptive trust must be evaluated in a complex way with realistic, scalable, and heterogeneous implementations [12], [14], [19], [20].

As such, it needs a thorough, and system level assessment of the challenges and opportunities this transition poses between 5G and 6G to inform research directions, work on standardization and practical deployment plans. The objective of this survey is to integrate knowledge of technical literature, architectural proposals, and application driven research in order to find out fundamental limitations of existing 5G systems, considerations of new enabling technologies that will empower 6G, and brief engineering and scientific considerations that are required to make 6G networks secure, scalable, sustainable, and capable of supporting future data based societies.

## III. RELATED WORK

Along with the vision of 6G technology, extensive research has been done across several fields ranging from network architecture and artificial intelligence to security and novel applications. This section reviews key literature pertaining to these subjects with a view to establishing the state of the art and existing challenges. The discussion is divided into five subjects: (A) Evolution from 5G to 6G, (B) Network Architecture and Performance Studies, (C) Artificial Intelligence and Machine Learning in 6G, (D) Security and Privacy in 6G Networks, and (E) Emerging Applications.

### A. *Evolution from 5G to 6G*

There are studies that have explored the evolution from 5G to 6G networks technologically. Salameh and Tarhuni [1] wrote the enormous range of challenges and technologies involved in 6G, whereas Rischke et al. [2] wrote 5G campus performance metrics for real-world deployments. Evolution from 5G mobile communications to 6G networks is a gigantic technology leap driven by more capacity, less delay, and more connectivity. Lin [3] considers the 3GPP releases that ushered in 5G and 6G, stressing advancements in spectrum use, edge intelligence, and extended reality support. Qidwai et al. [4] discuss larger trends in development and highlight the changing role of AI integration and terahertz communication technologies. Concurrently, Alam et al. [5] presents vehicular edge networks as an exemplary case of the evolution of 5G to 6G with emphasis on edge intelligence and green connectivity. Combined, these works outline how 5G architecture and performance limits have informed the research roadmap for next generations 6G systems. Much of the existing literature, nonetheless, still exists in theory and standardization with little empirical evidence of 6G-enabling technology being implemented in real testbeds. Following this progress, researchers have also looked at how architectural shapes must evolve to support these evolved functions.

### B. *Network Architecture and Performance Studies*

Network Architecture is the foundation of 6G's successful deployment. The Next Generation Mobile Networks Alliance [6] defines main requirements for 6G-enabled network architectures focusing on cloud-native cores, open interfaces, and green topologies. Khalid and Rehan [7] present architectural optimizations for end-to-end orchestration and scaling, and Jain [8] discusses the evolution from 5G's service-based architecture to 6G's integrated terrestrial–non-terrestrial ecosystems. Gkonis et al. [17] follows up on these findings with a survey of potential 6G architectures for the future, in which software-defined infrastructure and green networking are put forward as top design goals. Cumulatively, these works confirm that modular, flexible, and software-defined architectures will be central to scalable and sustainable 6G deployment. Many of the architectural proposals are conceptual, highlighting the necessity of experimental validation and cross-domain interoperability research for real-world 6G deployment. Beyond architecture, intelligence for future automation and efficiency is led chiefly by advances in artificial intelligence and machine learning.

### C. *Artificial Intelligence & Machine Learning in 6G*

It is expected that Artificial Intelligence (AI) and Machine Learning (ML) will be the core of 6G networks, which will enable self-organization, intelligent utilization of resources, and context-responsive services. Huang et al. [9] suggested an adaptive spectrum allocation model that was founded on AI, whereas the study by Zhang et al. [10] examine latency-aware federated learning frameworks for 6G networks. Chen et al. [11] research sensor-based AI architectures enhancing situational awareness and network energy efficiency. A companion paper by Alsharif et al. [18] provides a snapshot of AI embedding in communication systems, enumerating cross-layer intelligence open challenges and key opportunities. These papers illustrate how AI and ML will define the performance, adaptability, and autonomy of future wireless systems. Nevertheless, all the existing models adopt simulated frameworks, and none discuss the scalability and ethical constraints of using AI-based decision-making in large 6G networks. As networks become increasingly autonomous, these AI-based systems introduce new security challenges to secured and privacy-maintaining communication.

### D. *Security & Privacy in 6G Networks*

Security and privacy remain essential issues for 6G systems. Alsharif et al. [12] address the possible threats and suggest blockchain-based trust systems to distribute systems. Elayan et al. [13] consider AI-based anomaly detection and post-quantum cryptography, and Zhang and Yu [14] combine privacy-preserving protocols to ultra-dense environments. Saeed et al. [19] also elaborate such proceedings in an elaborate survey that labels 6G security threats on physical, connection, and service layers, but Mao [20] does not overlook edge computing to fight with privacy vulnerability. Taken together, these inputs illustrate the necessity of adaptive, lightweight, and quantum-resistant security paradigms of 6G networks. Nevertheless, regardless of these improvements, there are still questions of how to build cohesive models that dynamically evolve as hybrid attack surfaces in a large-scale, decentralized 6G network. Finally, the groundbreaking effect of 6G is attested via its growing real-world deployments across IoT, XR, and vehicular systems.

### E. *Emerging Applications*

New applications demonstrate the disruptive nature of 6G technologies. Liu et al. [15] study the 6G capabilities of serving the massive IoT connectivity via intelligent resource sharing, whereas Novak [16] explores the extended-reality (XR) and immersive communications. Alam et al. [5] show how vehicular edge computing is integrated with 6G potential, which allows achieving ultra-reliable, low-latency, and sustainable communication. Qadir et al. [21] provide a general overview of 6G-based applications and technologies in IoT, and Gallego-Madrid et al. [22] emphasize vehicular networks as one of the moving forces in designing the 6G system. Collectively, this research demonstrates that the effects of 6G go beyond enhanced speeds, which allow intelligent, immersive, and hyper-connected ecosystems in a variety of areas. However, most application studies are limited to domain-specific, which means that there is a knowledge gap in the cross-domain integration systems through which IoT, XR, and vehicular ecosystems are brought together into a unified 6G platform.

Collectively, the reviewed literature provides a solid foundation of 6G research in the field of technology, architecture, and application. Nonetheless, there are still consistent gaps in the integration of AI-controlled automation and scalable, risk-free, and empirically tested designs. Although the available literature offers good theoretical and conceptual foundation, additional studies are necessary to integrate these developments in real situations of deployment. These observations are developed further in the sections below by suggesting a combined strategy to overcome the efficiency, adaptability, and reliability problems of 6G networks.

| Category | Focus | Key Contribution | Year | Ref. |
|---|---|---|---|---|
| 5G→6G Evolution | Overview of 3GPP releases | Standardization and technology roadmap | 2025 | [3] |
| Architecture | Cloud-native, open interface design | 6G core scalability and flexibility | 2024–2025 | [6], [17] |
| AI/ML | Federated and adaptive learning | Latency reduction, network optimization | 2024–2025 | [9], [10], [18] |
| Security | Blockchain, quantum-safe mechanisms | Layered privacy and anomaly detection | 2023–2025 | [12], [19], [20] |
| Applications | IoT, XR, Vehicular integration | Domain-specific 6G use cases | 2023–2025 | [15]–[22] |

## IV. METHODOLOGY

A systematic literature analysis was conducted to evaluate the current state of 5G technologies, identifying their limitations and exploring proposed advancements that will guide the transition into 6G networks. The approach is structured around the systematic collection, assessment, comparison, and combination of existing literature. The objective being to organize current knowledge, highlight technological gaps, and present informed recommendations for next-generation wireless communications.

To support a clear and well-defined methodological structure, the organization of this analysis was informed by the framework proposed by Carrera-Rivera et al. Their guidelines provide a step-by-step model for planning and conducting literature-based research in computer science, including the development of review protocol, definition of research questions, creation of inclusion and exclusion criteria, quality assessment procedure and structured data-extraction methods. [51]

To begin, a literature search strategy was implemented using leading academic and technical databases, including IEEE Xplore and ACM Digital Library. These databases were chosen due to its relevance to telecommunications research and their reputations for reviewed publications. Using keywords and phrases like "5G limitations", "Challenges from 5G to 6G", and "6G architecture" produced a wide range of scholarly works.

The initial search yielded a broad collection of articles, which were then filtered using predefined inclusion and exclusion criteria. Only peer-reviewed journal articles, conference papers, and technical surveys published between 2020 and 2024 were included to maintain recency and relevance. Papers lacking technical depth and clear methodical foundations were excluded. After screening, the remaining set of publications used represented the most credible and relevant findings forming the basis for a deep analysis.

The analytical process in this paper was based on the relevant publications that were used throughout the paper. All papers added valuable technical knowledge, empirical research, and suggestions that would take a perfect picture of the transition between 5G and 6G. Other papers laid focus on physical-layer constraints like propagation loss, and energy inefficiency and others on architectural redesign and new application requirements.

The insights extracted from these publications were evaluated using a comparative and thematic approach. Each study was analyzed to identify its core objectives, the problems highlighted, and the innovations it proposed. Once the relevant literature was collected, all selected papers were organized and made into a table to facilitate structured analysis. This table included information such as publication year, authors, research focus, identified limitations and proposed solutions. Categorizing and tabulating the papers provided a clear visualization of how different studies get along, diverge and complement one another. The table specifically allowed the research to focus on individual aspects of the 5G-6G transition and made it possible to observe patterns across literature, to determine which issues were consistently emphasized by researchers and to identify areas where perspectives differed.

This precise comparison of studies enabled the development of findings that reflect collective academic consensus rather than relying on the viewpoint of one author. When multiple papers presenting similar limitations or proposed similar technical advancements it was treated as a strong requirement indicator for what needs to be done. Conversely when studies had differences it was analyzed to understand the underlying causes whether technical, methodological or conceptual. The analysis formed results that were grounded with academic knowledge ensuring that conclusions later in the paper are based in reliable and well-supported evidence.

The systematic literature was the most appropriate method for this research because the transitions from 5G to 6G remains largely in its conceptual and developmental stages, meaning that comprehensive experimental data and standardized implementations or even large-scale performance evaluations do not exist yet. As a result, the most reliable and academically rigorous way to understand the current technological landscape is through structured evaluation research. Unlike opinion-based reviews, the systematic approach ensures that every paper is screened using predefined criteria and that the insights are extracted consistently throughout all papers. Contributions of several independent studies are organized, compared, and synthesized, and this approach allows forming a conclusion based on the overall knowledge of the academic community, which is the most appropriate and reliable methodology on a topic where theoretical proposal and conceptual framework are more predominant than the full-fledged technologies or real-world implementations.

Overall, the methodological framework allowed turning the process of reviewing the scholarly materials into a transparent, structured, and credible endeavor. The comparative analysis conducted on the selected publications, their extraction of main contributions, and synthesis of their insights allowed the study to lay a solid ground on the existing limitation of 5G and the new technologies that are being introduced towards the 6G. With this framework in place the results and analysis derived the collective patterns, themes and research trends identified throughout the literature.

# V. RESULTS & ANALYSIS

The current section provides a comparative discussion of the reviewed literature on the migration between 5G and 6G networks based on essential topics that reappear in the technical, architectural, and security levels. Instead of the user-based empirical data, the results are obtained through the systematic comparison of the available 5G performance research, initial 6G proposals, and the cross-domain surveys which consider the spectrum usage, network architecture, AI integration, security frameworks and future applications. Articles like the general overview of Salameh and Tarhuni, where 5G and corresponding technologies are examined in terms of challenges and technologies in the transition to 6G [1], the empirical measurement of 5G campus deployments by Rischke et al. [2], the recent architectural and AI-native 6G surveys [3], [17], [23], [24], and [29], [30] all give the same image, 5G systems are robust, but structurally inadequate to support the proposed usages of 6G.

The analysis is organized into five subsections: (A) spectrum and physical-layer constraints, (B) architectural evolution and interoperability, (C) AI-driven network intelligence, (D) security and privacy, and (E) emerging applications and system-level opportunities. Each subsection provides a narrative synthesis followed by a concise results table summarizing the main comparative findings.

## *A. **Spectrum Constraints and Physical-Layer Limitations***

Across almost all the surveyed works, spectrum scarcity and propagation limitations arise as the most immediate technical constraints in current 5G systems and as primary drivers for 6G research. Salameh and Tarhuni [1] and Qidwai et al. [4] emphasize that 5G's current spectrum model, based mainly on sub-6 GHz and fragmented mid-band allocations; cannot sustain anticipated traffic growth driven by enhanced mobile broadband, massive IoT, and ultra-reliable low-latency applications. Empirical measurement work by Rischke et al. [2] confirms that in realistic 5G campus deployments, mid-band resources become congested under dense user and device conditions, leading to performance degradation that is not apparent from theoretical capacity estimates alone. Similar concerns about mid-band exhaustion and spectrum fragmentation are echoed in broader 6G surveys, which warn that even early 6G deployments are likely to inherit these pressure points if spectrum policy remains unchanged [23], [24], [27].

At the physical layer, 5G's use of mmWave frequencies is widely recognized as a double-edged sword. On the one hand, mmWave provides significantly higher peak data rates; on the other hand, measurements in realistic urban and indoor environments show that path loss, blockage, and penetration limitations drastically constrain coverage [1], [2], [4]. Studies on mmWave propagation document sharp attenuation through walls, glass, foliage, and even human bodies, as well as rapid signal degradation under rain and atmospheric absorption. These results support the conclusion that 5G's mmWave bands are well-suited for localized hotspots but are fundamentally limited as a primary carrier for ubiquitous, wide-area connectivity. To address these limitations, recent 6G research increasingly explores the concept of integrated sensing and communication (ISAC), in which radio signals are jointly used for data transmission and environmental sensing. Surveys by Liu et al. demonstrate that ISAC-enabled systems can improve spectrum efficiency by allowing communication and sensing functions to coexist within the same frequency bands, particularly at sub-THz and THz frequencies [43]. Xiao et al. further show that joint waveform and resource design can enhance situational awareness while reducing redundant spectrum usage [44]. These approaches suggest that 6G may partially mitigate spectrum scarcity not only through access to higher frequencies but also through more intelligent, multifunctional use of the radio medium itself.

6G literature responds to these constraints by proposing a shift towards sub-terahertz and terahertz (THz) bands, which promise terabit-per-second data rates and support for ultra-high-resolution sensing. Salameh and Tarhuni [1] and more recent surveys on THz communications highlight these bands as key enablers of 6G's ambitious performance targets [25], [26]. However, the same works underscore that propagation losses, hardware immaturity, and atmospheric absorption at these frequencies are even more severe than in mmWave. Jiang et al. survey THz propagation, device technology, and integrated sensing, concluding that without substantial advances in transceiver design, beamforming, and channel modeling, THz communications cannot be deployed at scale in realistic heterogeneous environments [25].

Future 6G studies therefore place strong emphasis on compensating mechanisms. Among the most extensively discussed compensating technologies are reconfigurable intelligent surfaces (RIS) that dynamically manipulate electromagnetic waves to enhance coverage, minimize interference, and circumvent high loss propagation at high frequencies. Basar et al. prove that RIS-aided communication can be very useful in improving link reliability in areas where line-of-sight is hindered and thus it is one of the most appealing systems to deploy THz systems [42]. In parallel, semantic communication paradigms have been proposed to further improve spectrum utilization by transmitting only task-relevant or meaning-oriented information rather than raw data streams [41]. By jointly leveraging RIS-enabled propagation control and semantic-aware transmission strategies, 6G networks may achieve substantial gains in efficiency despite harsh physical-layer constraints. Reconfigurable intelligent surfaces (RIS), ultra-massive MIMO arrays, and photonic transceivers themselves are repeatedly suggested as methods to control and enhance THz links, enhance coverage, and suppress interference [1], [11], [17], [25], [26]. Another common pattern is AI-assisted spectrum management; Huang et al. [9] and Zhang et al. [10] provide machine-learning-based allocation, prediction, and coordination using heterogeneous bands, and wider surveys believe AI-native spectrum control will be required to utilize fragmented, dynamic, and shared spectral environments [23], [29], [30].

Taken together, the surveyed papers show strong consensus on three key points. First, 5G's spectrum model, especially in the mid-band, is approaching saturation in high-demand markets. Second, 5G's physical-layer characteristics at mmWave frequencies limit its reliability and coverage in real deployments, making it unsuitable as the sole foundation for next-decade services. Third, 6G's proposed reliance on sub-THz and THz frequencies offers substantial performance gains but introduces more severe propagation, hardware, and energy challenges that the current technology ecosystem is not yet capable of solving.

### B. *Architectural Evolution and Interoperability Gaps*

From an architectural perspective, the reviewed literature confirms that the models of the current 5G core and radio access network are not fully aligned with the requirements of 6G's envisioned intelligent, hyper-connected ecosystems. The white paper on 6G network architecture by the NGMN Alliance [6], analysis by Khalid and Rehan on integrated terrestrial and non-terrestrial systems [7], and Jain's work on architectural evolution [8] all point to deficiencies within 5G's service-based architecture. Though the service-based architecture and cloud-native design certainly represent a big leap forward from previous generations, these works have shown that 5G relies heavily on centralized control and monolithic orchestration models, which become bottlenecks in ultra-dense, highly dynamic environments.

More recent surveys on 6G architectures reinforce this perspective. Gkonis et al. [17] emphasize that future 6G networks must be modular, software-defined, and capable of integrating terrestrial networks with satellite, aerial, and maritime systems. Similar visions appear in Akbar's overview of 6G challenges and enablers [23] and Amgoune and Mazri's discussion of requirements and technical specifications for the 5G–6G evolution [27]. These works always refer to 6G as a network of networks, with cloud, edge, and device-level intelligence working together and non-terrestrial networks (NTNs) being inherently integrated instead of being considered an overlay. They, however, also point out that pragmatic architectures that can coordinate these heterogeneous domains are still quite theoretical and that standardization has not reached a very high level yet. It is hoped that one of the key solutions to these architectural gaps will be non-terrestrial networks (NTNs), which will be able to connect the different areas beyond the scope of ground-based infrastructure. According to Giordani et al, low Earth orbit (LEO) satellites constellations, high-altitude platforms, and unmanned aerial systems may offer global coverage, resilience, and continuity of service to 6G networks in the future [45]. Kodheli et al. further argue that integrating satellite communications into the core architecture enables seamless support for mobility-intensive and remote-area applications that remain challenging for terrestrial-only systems [46]. Despite their promise, these NTN-integrated architectures raise complex challenges related to latency management, handover coordination, and cross-domain interoperability, which remain open research problems in current 6G proposals.

Interoperability emerges as a recurring challenge. Although 5G introduces features like network slicing and service-based interfaces, the surveyed literature demonstrates that cross-domain interoperability between different operators, vendors, and vertical sectors is limited. Campus and private 5G deployments studied by Rischke et al. [2] show how tailored architectures can deliver strong local performance, yet they remain isolated and difficult to interconnect with broader public networks, let alone with future NTNs. In contrast, many 6G papers envision architectures that seamlessly integrate terrestrial access, high-altitude platforms, UAV-assisted relays, and LEO satellite constellations [6], [7], [17], [23], [28], [29]. Despite this, few works provide concrete mechanisms for harmonizing mobility management, routing, and resource allocation across these layers under realistic load conditions.

Architectural surveys also converge on the idea that 6G must be AI-native rather than merely AI-assisted. This is reflected both in AI-focused works [9]– [11], [18] and broader 6G architectural reviews [23], [29], [30]. Edge computing will shift from an optional add-on in 5G to a core enabler of real-time inference, local decision-making, and distributed learning across geographically dispersed nodes in 6G. In the words of Ogenyi et al., this translates to a shift from "data-centric" to "intelligence-native" architectures, whereby meaning, context, and adaptivity are central design parameters rather than afterthoughts [30].

Although these converging visions exist, the comparison analysis indicates that there exists a large gap between high-level architectural proposals and proven, large-scale prototypes. Many surveyed works make use of analytical models or simulations; few offer evidence-of-concept testbeds, and these are usually of isolated subsystems, and not of complete architectures based on 6G systems. This gap therefore indicates that though there is agreement on the way architectural evolution should be a lot of engineering and standardization still needs to be done before 6G frameworks can be converted into deployable infrastructures.

### C. *AI-Driven Network Intelligence: Capabilities and Risks*

A particularly strong theme across the surveyed literature is the central role of artificial intelligence (AI) and machine learning (ML) in defining 6G network behavior. In 5G systems, AI is largely used for localized optimization, such as traffic prediction, scheduling, and parameter tuning, while core control logic remains mostly deterministic. Studies by Huang et al. [9], Zhang et al. [10], and Chen et al. [11] demonstrate that even in 5G-advanced and early 6G scenarios, AI can significantly improve spectrum utilization, energy efficiency, and latency through adaptive resource management and smart sensing architectures.

However, the conceptual leap proposed for 6G goes much further. Alsharif et al. and multiple recent surveys describe 6G as inherently AI-native: network functions such as spectrum allocation, mobility management, handover decision-making, and anomaly detection are expected to be driven by distributed learning models operating at the core, edge, and device levels [23], [24], [29], [30]. Ogenyi et al. say that 6G will shift from traditional data-pipe models to semantic and goal-oriented communications, where AI drives not only how data is transmitted but also what information is needed to meet application-level objectives [30]. As AI-driven control becomes deeply embedded in 6G architectures, concerns regarding trust, transparency, and accountability grow increasingly significant. Zhang et al. emphasize that trustworthy AI frameworks must be incorporated into network design to ensure explainability, robustness, and fairness in automated decision-making processes [47]. Similarly, Boutaba et al. highlight that while AI enables unprecedented levels of autonomy and optimization in networking, it also introduces risks related to model bias, adversarial manipulation, and unpredictable behavior under dynamic conditions [48]. These findings suggest that AI-native 6G networks must balance performance gains with safeguards that ensure reliable and interpretable operation, particularly in safety-critical and mission-sensitive environments.

At the same time, the literature reflects increasing awareness that AI-native control introduces new vulnerabilities and design trade-offs. Several works observe that most proposed AI models for network control are only evaluated under simplified assumptions and simulated environments, while paying little attention to adversarial behavior, distribution shifts, or real-time constraints [9, 10, 18, 23, 30]. Complementary security-focused surveys show that ML models are vulnerable to adversarial examples, poisoning attacks, and inference-based privacy breaches that could be leveraged to manipulate network behavior, disrupt key services, or exfiltrate sensitive information [31, 35].

Federated learning and edge intelligence, recently publicized as approaches that offer greater protection of privacy compared to centralized training [10], also emerge as double-edged solutions: while they reduce the need to collect raw data in a central location, they introduce more attack surfaces and coordination challenges. Heterogeneous device capabilities, intermittent connectivity, and participants prone to unreliability complicate model convergence and robustness. This is consistent with the broader set of open concerns related to robustness, explainability, governance, and energy overhead that have been emphasized in general AI-native 6G surveys [23], [29], [30], [34].

All in all, the literature is united in the opinion that AI is necessary to handle the scale, complexity, and real-time demands of 6G networks, and modern AI frameworks and assessment tools are not yet sufficient to support safety-critical large-scale implementation. This is the main tension where AI is at the center of the opportunities and the threats of the 5G-6G transition.

### D. *Security, Privacy, and Trust in 5G-6G Evolution*

Security and privacy are repeatedly identified in the literature as some of the most critical and least mature aspects of the 5G–6G transition. In your existing references, Alsharif et al. [12], Elayan et al. [13], Zhang and Yu [14], Saeed et al. [19], and Mao [20] present a layered view of vulnerabilities from the physical, connection, and service layers, with a focus on edge computing, IoT, and blockchain-based trust frameworks. Collectively, they argue that while 5G brings stronger cryptographic mechanisms and network slicing-based isolation, it also has amplified exposure due to expanded attack surfaces, virtualized network functions, and densified infrastructure.

Further 6G-focused surveys give more depth to this perspective: Wang et al. present one of the earliest comprehensive treatments on 6G security and privacy, identifying new threat classes associated with AI-native control, open interfaces, and integration with quantum computing [32], while Hakeem et al. discuss 6G security requirements in terms of confidentiality, integrity, availability, and trust, emphasizing that traditional security-by-addition approaches are infeasible in resource-constrained and latency-sensitive 6G settings [33]. More recent surveys, including Saeed et al.'s multi-layer 6G security analysis [19][38], systematically categorize threats across physical, connection, and service layers and propose adaptive, cross-layer defenses combining cryptography, physical-layer security, and AI-based anomaly detection.

Privacy becomes an increasingly challenging issue. The authors of [31] survey the privacy of personal and non-personal data in B5G/6G settings, underlining that emerging applications, such as pervasive sensing, digital twins, and vehicular data sharing, will blur the boundary between personal and contextual data and expose users to new forms of profiling and inference. This concern goes in line with works such as the one by Elayan et al. on edge intelligence and post-quantum cryptography [13] and Zhang and Yu on quantum-safe protocols [14], which advocate for making privacy-preserving mechanisms native to 6G architectures rather than optional features.

Security and privacy surveys also increasingly address the role of AI in both defending and attacking 6G networks. On one hand, AI can be used to detect anomalous traffic, identify misbehaving nodes, and adaptively reconfigure network resources in response to threats [13], [19], [20], [35]. On the other hand, AI models themselves become targets, susceptible to extraction, poisoning, and adversarial manipulation. The "From 5G to 6G: Security, Privacy, and Standardization Pathways" survey emphasizes that emerging technologies such as network softwarization, virtualization, and distributed ledgers introduce new vulnerabilities that must be anticipated during standardization rather than patched later [35].

All in all, in the comparative analysis, 5G has already posed some serious security and privacy concerns, but 6G provides much more resources to attack (AI, THz sensing, NTNs, and massive-scale IoT). There is agreement in the literature that security and privacy should be made design primitives in 6G and not peripheral issues, although specific and unified models of how to accomplish this are still largely in development.

### *E. Emerging Applications, Sustainability, and System-Level Opportunities*

The final major theme, which cuts across the surveyed literature, involves the applications and system-level opportunities that are driving migration from 5G toward 6G. Indeed, your existing references are quite diverse, ranging from vehicular edge networks [5] to IoT ecosystems [15], extended reality (XR) services [16], and vehicular-driven aspects in the design of 6G [21], [22]. Generally, they indicate that although 5G can allow the early adoption of these applications, some of its drawbacks such as latency, reliability, scalability, and integration limit it to the potential of such applications.

The recent 6G surveys, including those by Mohsan et al. [24], Giordani et al. [28], and Akbar [23] are a continuation of this discussion, highlighting future directions of holographic tele-presence, scaled digital twins, and genuinely immersive XR systems, where haptic responses and semantic interactions are needed. These applications require not only increased peak rates, but also deterministic end-to-end latency, ultra-reliable links, and joint optimization of communication and computing and sensing resources, and performance that is beyond the capabilities of current 5G deployments, particularly in wide-area and mobility-intensive applications.

Vehicular and transportation systems receive particular attention in the surveyed literature. Alam et al. [5] and Gallego-Madrid et al. [22] illustrate how vehicular applications—such as cooperative driving, real-time HD mapping, and multi-access edge-assisted perception, impose strict constraints on latency, availability, and handover performance, highlighting 5G's difficulties in providing seamless coverage under high mobility. 6G-oriented works propose tighter coupling among vehicular nodes, roadside units, edge servers, and NTNs to achieve global, resilient vehicular connectivity [21]– [23], [28].

Another interesting theme that has picked up in more recent 6G surveys is that of sustainability. While earlier generations optimized basically for capacity and coverage, newer works argue energy efficiency and environmental impact need to become first-class design considerations for 6G. Akbar's challenge analysis [23] and Kamran et al.'s research on energy-aware 6G network design [34] stress that AI-native architectures, THz hardware and dense cell deployments threaten to increase energy consumption significantly unless carefully managed. Energy harvesting, renewable integration, and AI-based energy optimization are offered as mitigation strategies but are very theoretical. Recent surveys on green and sustainable 6G networking emphasize that energy efficiency must be treated as a core design objective rather than a secondary optimization. Zhang et al. argue that without careful architectural planning, dense deployments, AI-native control, and high-frequency transceivers could significantly increase the carbon footprint of future networks [49]. Kamran et al. also suggest power management based on energy awareness, adaptive sleep schedules, and AI-based power management as potentially viable compromises between the needs of performance and sustainability [34]. Nevertheless, these techniques are mostly conceptual and experimental validation and testbeds are necessary to test their applicability on a large scale.

The literature analysis thus indicates that the most radical applications of 6G; immersive XR, digital twins, pervasive IoT, autonomous systems, and globally available critical communications are closely linked to sustainability, security and architectural issues, but as it currently stands most such solutions exist as a concept in a lab or on a few pilot projects.

## VI. DISCUSSION

The findings of this study indicate that the transition from 5G to 6G is not a simple linear upgrade in speed or capacity, but a systemic transformation that touches every layer of network design, from spectrum usage and physical-layer technologies to architecture, intelligence, and security. The research consistently shows that while 5G has successfully enabled enhanced mobile broadband, IoT connectivity, and low-latency services, its dependence on fragmented spectrum, mid-band congestion, and context-limited mmWave deployment makes it inadequate for emerging 6G use cases such as holographic communications, large-scale digital twins, and prevalent autonomous systems [1]–[5], [23], [28]. At the same time, many of the proposed 6G enablers, such as THz communications, reconfigurable intelligent surfaces, ultra-massive MIMO, and photonic transceivers, are still immature, revealing a gap between theoretical potential and deployable solutions [11], [17], [25], [26]. This indicates that substantial work remains before practical deployments can be realized.

A central implication of the analysis is that 6G will require a fundamental architectural shift rather than periodic extensions of 5G. The move toward a "network of networks" that integrates terrestrial, aerial, and satellite domains, combined with hierarchical intelligent computing, introduces powerful opportunities for ubiquitous coverage and context-aware services, but also amplifies orchestration and interoperability challenges [6]-[8], [17], [23], [27]. Current 5G deployments already struggle to provide seamless integration capabilities across vendors, operators, and private/public domains; extending this model to natively include non-terrestrial networks and different industry sectors makes the complexity even greater. The strategy to address this complexity can be seen in the vision of AI-native architectures, in which the learning models determine spectrum allocation, mobility, routing, and anomaly detection, which, however, also creates new failure modes, such as model drift, adversarial manipulation, and unclear decision making in situations with safety criticality [9], [10], [18], [29], [30]. Standardization and regulatory frameworks coordination will also be necessary in the achievement of this vision. The ITU's IMT-2030 initiative and ETSI's emerging 6G white papers emphasize that future networks must be globally interoperable, secure by design, and adaptable to regional policy constraints [36], [38]. Al-Mousa et al. also emphasize the fact that the delayed alignment between technological innovation and standardization has historically contributed to dispersed installations and security breaches, which is increased in AI-driven and software-defined 6G systems [35]. These findings suggest that regulators, industry alliances, policymakers, and researchers need to work together at the initial stage of the design to make sure that 6G systems develop in a consistent manner instead of making regionalized choices that are mutually incompatible.

Security, privacy, and trust emerge from the synthesis as crosscutting constraints rather than isolated design considerations. Compared to 5G, 6G is expected to operate with denser IoT deployments, richer sensing-including sensing-capable THz links-and more autonomous decision making at the edge-all of which increase the attack surface across physical, connection, and service layers [12]- [14, 19, 20, 31-33, 35]. While blockchain-based authentication, post-quantum cryptography, and AI-driven anomaly detection are being proposed, the analysis indicates that most solutions have been evaluated in narrow, domain-specific settings and mostly lack validation under heterogeneous large-scale deployments. Hence, truly secure and privacy-preserving 6G networks will likely require integrated, multi-layer security-by-design approaches, co-evolving with standardization efforts rather than being retrofitted post-deployment [19, 32, 35].

Another significant implication of this work is that sustainability and system-level optimization need to be treated as first-order objectives. Many of the same technologies that enable performance at 6G, dense deployments, THz hardware, and AI-native control, risk increasing energy consumption and operational complexity if not holistically managed [23], [24], [34]. The literature indicates energy-aware architectures, green topology design, and AI-based energy optimization as well-motivated directions, but empirical evidence remains scant. This is in its turn speaking to the larger limitation of the existing research environment and, by proxy, of this survey, where most of the findings are the result of simulations, modeling and early testbeds, instead of full-fledged, real-world 6G experiments. Consequently, even though this research has solidified the technical, architectural, and security priorities of the transition between 5G and 6G, future research should be supplemented by large-scale experimental validation, extensive inter-industry integration projects, and regulatory and socio-ethical research. It is only in these conditions that the opportunities of 6G, intelligence-native, immersive, secure, and sustainable connectivity, can be achieved without recreating or exerting magnification over the constraints already being encountered at the preceding phases of the development of wireless networks.

## VII. CONCLUSION & FUTURE WORK

The transition between 5G and 6G extends beyond improvements in speed; it represents a paradigm shift of how the communication systems will be in future. Due to this study, it was revealed that despite the significant improvements made by 5G such as reduced latency, higher data rates, and IoT devices, 5G has several limitations. Such constraints include restricted spectrum availability, signal attenuation at high frequencies, and the challenges of supporting extremely large numbers of devices. Technologies are expanding particularly in such aspects as virtual reality, autonomous vehicles, smart cities, and AI; hence the network should be adapted to such changes.

6G is being developed to solve these issues and introduce new possibilities. 6G is anticipated to operate under much higher frequencies such as terahertz waves, be very high speed, have the capability of real-time sensing, and incorporate AI in the network. Nevertheless, a lot of these technologies are yet to be developed. Terahertz communication hardware is immature; signals, being short-lived at high frequencies, are not only experimental but new devices such as reconfigurable intelligent surfaces are in development. This implies that 6G is a highly promising technology, but it requires numerous engineering issues to be resolved before it can be a reality.

In 6G, AI will play a much larger role than it was in any other generation. AI will not only assist the network but will also contribute to its control, making decisions regarding the spectrum, routing, mobility, and even security. These positive results such as smarter traffic control and quicker reaction come at a price, as there are new risks brought up. Any attack of the AI model, its manipulation, or errors may impact the whole of the network. As a result, future networks will need strong protections to make sure AI-powered decisions are safe, fair, and reliable.

Security and privacy are some of the biggest concerns as we move toward with 6G. With millions of new IoT devices, new sensing technologies, and so much more data being shared across devices, the risk of cyberattacks will certainly increase. Researchers suggest using tools like blockchain, quantum-safe encryption, and AI-based threat detection, but these ideas need more testing. The main challenge will be designing a network that is fast and intelligent while still protecting users' data and ensuring privacy.

Future research should therefore prioritize large-scale experimental validation to bridge the gap between theoretical models and deployable 6G systems. This includes the development of real-world testbeds for terahertz communication, AI-native control frameworks, and integrated terrestrial–non-terrestrial architectures, as well as the evaluation of governance mechanisms for trustworthy AI operation [36], [37]. Additionally, spectrum policy evolution and international coordination will play a decisive role in determining how effectively 6G technologies can be deployed on a global scale [50]. Interdisciplinary collaboration will be key to the achievement of the full potential of next-generation wireless networks by addressing these challenges. Finally, 6G will allow applications that are not supported by 5G. This will include holographic communication, full-immersion XR, and autonomous transportation systems. These new features also require better energy efficiency and sustainable designs, so networks don't consume too much power. Further studies ought to be based on practical experiments, as the majority of the 6G concepts are not yet practical. To enhance the safety and transparency of AI, as well as to establish a common set of global standards in which satellites, drones, and earth networks may communicate, terahertz communication requires large testbeds; to design energy-efficient hardware that could achieve dense 6G settings. Better and versatile security systems should also be developed to secure all levels of these highly interconnected systems. While the shift from 5G to 6G presents many challenges, it offers a major opportunity to create communication networks that are smarter, safer, and more advanced than ever before.

## REFERENCES

[1] A. Salameh and M. Tarhuni, "From 5G to 6G—Challenges, Technologies, and Applications," *Future Internet*, vol. 14, no. 4, p. 117, Apr. 2022.

[2] J. Rischke, P. Sossalla, S. Itting, F. H. P. Fitzek, and M. Reisslein, "5G Campus Networks: A First Measurement Study," *IEEE Access*, vol. 9, pp. 121786–121803, 2021.

[3] X. Lin, "3GPP Evolution from 5G to 6G: A 10-Year Retrospective," *Telecom*, vol. 6, no. 2, p. 32, 2025.

[4] M. A. Qidwai, M. A. Siddiqui, and M. S. Qidwai, "Trends in the Development of 5G and 6G Communication Systems," *Automatic Documentation and Mathematical Linguistics*, vol. 57, no. 5, pp. 247–258, 2023.

[5] A. Alam, M. A. Rahman, M. S. Hossain, and G. Muhammad, "Edge Intelligence-Empowered Vehicular Networks Toward 6G," *IEEE Transactions on Intelligent Transportation Systems*, vol. 24, no. 10, pp. 10232–10243, Oct. 2023.

[6] NGMN Alliance, *Network Architecture Evolution Towards 6G*, White Paper v1.0, Feb. 2025.

[7] M. S. Khalid and A. Rehan, "6G Network Architecture: Toward Integrated Terrestrial and Non-Terrestrial Systems," *arXiv preprint arXiv:2411.18836*, 2024.

[8] R. Jain, "Evolution of Network Architecture Towards 6G," *CSE 574 Lecture Notes*, Washington University in St. Louis, 2024.

[9] K. Huang, X. Zhou, and Y. Xu, "AI-Driven Spectrum Management for 6G Wireless Networks," *IEEE Communications Magazine*, vol. 63, no. 7, pp. 42–49, 2024.

[10] Y. Zhang, C. Li, and T. Han, "Federated Learning for 6G Networks: Architecture, Challenges, and Future Directions," *Science China Information Sciences*, vol. 67, no. 8, p. 4337, 2024.

[11] L. Chen, X. Wu, and P. Zhou, "Smart Sensing and AI Architectures for 6G Systems," *Sensors*, vol. 24, no. 6, p. 1888, 2024.

[12] M. Alsharif, A. Yadav, and J. Kim, "Secure Blockchain-Based Framework for 6G Communication," *IEEE Access*, vol. 13, pp. 18732–18745, 2025.

[13] E. Elayan, M. Erol-Kantarci, and R. Li, "6G Security: Challenges, Threats, and AI Solutions," *arXiv preprint arXiv:2108.11861v2*, 2024.

[14] H. Zhang and P. Yu, "Privacy Preservation in 6G Networks Using Quantum-Safe Protocols," *Telecommunication Systems*, vol. 89, pp. 233–247, 2025.

[15] X. Liu, Y. Wang, and J. Cheng, "6G-Enabled IoT: Massive Connectivity and Intelligent Resource Allocation," *IEEE Internet of Things Journal*, vol. 11, no. 3, pp. 2173–2187, 2025.

[16] T. Novak, "Extended Reality Applications in 6G Networks," *SET Journal of Emerging Technologies*, vol. 5, no. 1, pp. 45–52, 2025.

[17] P. K. Gkonis, A. Papathanassiou, and M. Kountouris, "A Survey on Architectural Approaches for 6G Networks," *Electronics*, vol. 6, no. 2, p. 27, 2025.

[18] A. Alsharif, F. Zhang, and Y. Kim, "Overview of AI and Communication for 6G Networks: Fundamentals, Challenges, and Future Opportunities," *Science China Information Sciences*, vol. 68, no. 3, article 171301, 2025.

[19] M. M. Saeed, R. U. Khan, and S. Yousaf, "A Comprehensive Survey on 6G Security: Physical, Connection, and Service Layers," *Journal of Network and Systems Management*, vol. 33, no. 1, 2025.

[20] B. Mao, "Security and Privacy on 6G Network Edge: A Survey," *IEEE Communications Surveys & Tutorials*, vol. 25, no. 2, pp. 1550–1578, 2023.

[21] Z. Qadir, M. Tariq, and A. Khan, "Towards 6G Internet of Things: Recent Advances, Use Cases, and Technologies," *Journal of Network and Computer Applications*, vol. 223, p. 103620, 2023.

[22] J. Gallego-Madrid, R. Perez-Jimenez, and C. Gomez, "The Role of Vehicular Applications in the Design of Future 6G Networks," *Vehicular Communications*, vol. 41, p. 100650, 2023.

[23] M. S. Akbar *et al*., "On Challenges of Sixth-Generation (6G) Wireless Networks," *J. Netw. Comput. Appl.*, vol. 235, 2025, Art. no. 104040.

[24] S. A. H. Mohsan *et al*., "A Contemporary Survey on 6G Wireless Networks: Architecture, Technologies, and Challenges," *arXiv preprint* arXiv:2306.08265, 2023.

[25] W. Jiang *et al*., "Terahertz Communications and Sensing for 6G and Beyond: A Comprehensive Survey," *IEEE Commun. Surveys Tuts.*, vol. 26, no. 1, pp. 1–55, 2024.

[26] I. F. Akyildiz, A. Kak, and S. Nie, "6G and beyond: The future of wireless communications systems," *IEEE Access*, vol. 8, pp. 133995–134030, 2020.

[27] H. Amgoune and T. Mazri, "From 5G to 6G: Requirements, Challenges, and Technical Specification," *J. Adv. Informatics*, vol. 5, no. 4, 2024.

[28] M. Giordani *et al*., "Toward the 6G Network Era: Opportunities and Challenges," *IEEE Commun. Mag.*, vol. 58, no. 3, pp. 55–61, Mar. 2020.

[29] J. Hoydis *et al*., "Five Facets of 6G: Research Challenges and Opportunities," *ACM Comput. Surv.*, vol. 55, no. 9, pp. 1–37, 2023.

[30] F. C. Ogenyi *et al*., "A Comprehensive Review of AI-Native 6G: Integrating Semantic Communication, RIS, and Edge Intelligence," *Front. Commun. Netw.*, vol. 6, Art. 1655410, 2025.

[31] C. Sandeepa *et al*., "A Survey on Privacy of Personal and Non-Personal Data in B5G/6G Networks," *ACM Comput. Surv.*, 2024.

[32] M. Wang *et al*., "Security and Privacy in 6G Networks: New Areas and Challenges," *Digital Commun. Netw.*, vol. 6, no. 3, pp. 281–291, 2020.

[33] S. A. A. Hakeem *et al*., "6G Security Requirements and Challenges: A Comprehensive Survey," *Sensors*, vol. 22, no. 9, 2022, Art. no. 3412.

[34] R. Kamran *et al*., "Energy-Aware 6G Network Design: A Survey," *arXiv preprint* arXiv:2509.11289, 2025.

[35] A. Al-Mousa *et al*., "From 5G to 6G: A Survey on Security, Privacy, and Standardization Pathways," *arXiv preprint* arXiv:2410.21986, 2024.

[36] ITU-R, *IMT-2030 Framework for the Development of 6G Systems*, International Telecommunication Union, Geneva, Switzerland, Rep. ITU-R M.2160-0, 2023.

[37] 3GPP, *Study on Requirements for Next Generation Radio Access Technology*, TR 38.913, v18.0.0, 3rd Generation Partnership Project, 2024.

[38] ETSI, *6G White Paper: From Vision to Reality*, European Telecommunications Standards Institute, Sophia Antipolis, France, 2024.

[39] NGMN Alliance, *6G Drivers and Requirements*, White Paper, NGMN Ltd., Frankfurt, Germany, 2023.

[40] D. M. Strinati *et al.*, "6G: The next frontier," *IEEE Vehicular Technology Magazine*, vol. 16, no. 3, pp. 30–42, Sep. 2021.

[41] Y. Liu, Z. Qin, and M. Xiao, "Semantic communications for 6G: Vision, theories, and challenges," *IEEE Communications Magazine*, vol. 60, no. 1, pp. 86–92, Jan. 2022.

[42] E. Basar, "Reconfigurable intelligent surfaces for beyond 5G/6G wireless communications," *IEEE Journal on Selected Areas in Communications*, vol. 38, no. 11, pp. 2637–2653, Nov. 2020.

[43] F. Liu *et al.*, "Integrated sensing and communications: Toward dual-functional wireless networks for 6G and beyond," *IEEE Signal Processing Magazine*, vol. 39, no. 6, pp. 107–121, Nov. 2022.

[44] Z. Xiao *et al.*, "Joint communication and sensing for 6G networks," *IEEE Wireless Communications*, vol. 30, no. 1, pp. 44–51, Feb. 2023.

[45] M. Giordani *et al.*, "Non-terrestrial networks in the 6G era: Challenges and opportunities," *IEEE Network*, vol. 36, no. 2, pp. 22–29, Mar.–Apr. 2022.

[46] O. Kodheli *et al.*, "Satellite communications in the new space era: A survey and future challenges," *IEEE Communications Magazine*, vol. 59, no. 8, pp. 70–76, Aug. 2021.

[47] X. Zhang, Y. Mao, and K. Letaief, "Trustworthy artificial intelligence for 6G networks," *IEEE Network*, vol. 38, no. 1, pp. 112–118, Jan.–Feb. 2024.

[48] R. Boutaba *et al.*, "A comprehensive survey on machine learning for networking: Evolution, applications and research opportunities," *IEEE Journal on Selected Areas in Communications*, vol. 36, no. 10, pp. 2192–2217, Oct. 2018.

[49] Y. Zhang, S. Wang, and X. Chen, "Green 6G networks: Architecture, technologies, and challenges," *IEEE Communications Surveys & Tutorials*, vol. 25, no. 2, pp. 1091–1124, Second Quarter 2023.

[50] W. Saad, M. Bennis, and M. Chen, "A vision of 6G wireless systems: Applications, trends, technologies, and open research problems," *IEEE Communications Magazine*, vol. 58, no. 9, pp. 74–80, Sep. 2020.

[51] J. A. Carrera-Rivera, J. L. Enríquez, and R. Ramírez-Hernández, "Methodology for systematic literature reviews in computer science," *Information and Software Technology*, vol. 136, p. 106561, Aug. 2021.